\documentclass[]{spie} 

\usepackage[]{graphicx}

\title{Plasmons in nanoscale metal junctions: optical rectification and thermometry} 

\author{Douglas Natelson\supit{a}, Daniel R. Ward\supit{a}, Falco H{\"u}ser\supit{b}, Fabian Pauly\supit{c}, Juan Carlos Cuevas\supit{d}, David A. Corley\supit{e}, and James M. Tour\supit{e}
\skiplinehalf
\supit{a}Department of Physics and Astronomy, Rice University, MS 61, 6100 Main St., Houston, TX 77005, USA; \\
\supit{b}Institut f{\"u}r Theoretische Festk{\"o}rperphysik, Karlsruher Institut f{\"u}r Technologie (Universit{\"a}t), Wolfgang-Gaede-Str. 1, D-76131 Karlsruhe, Germany;\\
\supit{c}Institut f{\"u}r Theoretische Festk{\"o}rperphysik and DFG Center for Functional Nanostructures, Karlsruher Institut f{\"u}r Technologie (Universit{\"a}t), Wolfgang-Gaede-Str. 1, D-76131 Karlsruhe, Germany;\\
\supit{d}Departamento de F{\'i}sica Te{\'o}rica de la Materia Condensada, Universidad Aut{\'o}noma de Madrid, E-28049 Madrid, Spain;\\
\supit{e}Department of Chemistry, Rice University, MS 60, 6100 Main St., Houston, TX 77005, USA.
}

\authorinfo{Further author information: (Send correspondence to D.N.)\\D.N.: E-mail: natelson@rice.edu, Telephone: 1 713 348 3214}

\begin{document} 
  \maketitle

\begin{abstract}
We use simultaneous electronic transport and optical characterization measurements to reveal new information about electronic and optical processes in nanoscale junctions fabricated by electromigration.   Comparing electronic tunneling and photocurrents allows us to infer the optical frequency potential difference produced by the plasmon response of the junction.  Together with the measured tunneling conductance, we can then determine the locally enhanced electric field within the junction.  In similar structures containing molecules, anti-Stokes and Stokes Raman emission allow us to infer the effective local vibrational and electronic temperatures as a function of DC current, examining heating and dissipation on the nanometer scale.
\end{abstract}


\section{INTRODUCTION}


Plasmons are the collective modes of the conduction electron fluid in metal structures.  In the last decade, the ability to design and construct metal nanostructures on scales far smaller than the wavelength of visible light has led to the rapid growth of plasmonics, the engineering of plasmons\cite{Ozbay:2006,Schuller:2010}.  Plasmon excitations often couple well to electromagnetic radiation, and are accompanied by local evanescent fields that may be greatly enhanced relative to those of free-space incident or radiated waves.  These features enable plasmonic structures to function as nanophotonic antennas\cite{Muhlschlegel:2005,Yu:2007}, substrates for ``surface-enhanced'' spectroscopies\cite{Otto:1992,Hartstein:1980}, and components of novel optoelectronic devices\cite{Noginov:2009,Oulton:2009}.  


Metal nanoparticles are one traditional platform for plasmonics experiments.  These structures, often small in all dimensions relative to the wavelength of incident light, have a discrete spectrum of plasmon modes due to such confinement.  Great progress has been made in the quantitative understanding of such modes.  Given the dielectric function of the metal and the geometry of the nanoparticle, the interaction of a nanoparticle with incident radiation (the scattering, absorption, and the spatial distribution of the near field) may be calculated in detail, from analytical approaches (Mie theory) or numerical methods (finite element (FE) or finite difference time domain (FDTD)), depending on the symmetry of the system.  

Coupling plasmonic systems in the near field results in a hybrid system with a new set of normal modes.  This concept of plasmon hybridization\cite{Prodan:2003} has been very useful in understanding the complexities that may arise in such coupled systems.  For example, this is one way to think about the large field enhancements predicted in chains of self-similar plasmonic nanoparticles\cite{Li:2003}.  Nanoparticles aside, hybridization also conceptually explains the formation of localized plasmon modes when extended objects are placed in close proximity.  For example, a sharp metal tip brought into close proximity to a metallic ground plane leads to the formation of a tip-induced plasmon.  This mode arises from the hybridization of the continuum surface plasmon modes of both the ground plane and the extended metal tip\cite{GM:2011}.  Tip-induced plasmons have proven useful for tip-enhanced Raman spectroscopy\cite{Stockle:2000} and apertureless near-field scanning optical microscopy\cite{Wessel:1985}, and are directly relevant to light emission in scanning tunneling microscopy.  

Over the same period of time, planar molecular-scale junctions between electrodes have proven their worth as tools for examining electronic transport at the nanoscale.  With the development of controlled electromigration\cite{Park:1999} as a fabrication technique, it has become possible to create extended metal source and drain electrodes separated at the single nanometer scale.  Because of the exponential dependence of tunneling conduction on distance, the interelectrode tunneling conductance in these structures is dominated by a molecule-scale volume.  Forming such nanogaps in the presence of molecules (e.g., as a self-assembled monolayer on the metal electrodes) has been used extensively to create single-molecule transistors\cite{Park:2000,Park:2002,Liang:2002,Kubatkin:2003,Yu:2004,Natelson:2006,Song:2009}.  Electromigrated junctions in this case most commonly contain zero, one, or occasionally two molecules, though from electronic transport alone it can be difficult to assess this.

Electronic transport in nanoscale junctions involves rich physics and subtle issues, both practical and fundamental.  One practical concern has already been mentioned.  Every planar junction is different at the atomic scale, and there are no microscopy tools available to assess, e.g., the atomic-scale arrangement of metal atoms at the electrode tips, or the orientation and bonding of any molecules present.  On a more fundamental level, when a voltage bias is applied between the electrodes, causing electrons to flow from the source to the drain by one of several possible conduction mechanisms, how and where does dissipation take place?  The injection of hot carriers from the source drives the drain to have a nonthermal electronic distribution function on distance scales comparable to the inelastic mean free path for electrons in the metal.  Coupling between sufficiently energetic tunneling electrons and the local vibrational modes of the molecule can similarly drive the vibrational populations far from their thermal equilibrium values.  These couplings and the relaxation rates of local modes to the bulk phonons should depend strongly on molecular orientation and environment.  From transport measurements alone, it is extremely difficult to access this kind of microscopic information that would give insights into dissipative processes.


In recent years, we have found that nanoscale gaps between Au electrodes, prepared via electromigration, support plasmon modes localized to the nanogap, analogous to the tip-induced modes described above\cite{Ward:2007,Ward:2008,Ward:2008b}.  These modes were revealed through surface-enhanced Raman scattering (SERS) experiments with near-infrared illumination.  In this paper we review the recent progress in understanding these plasmon modes, as well as the insights into the physics of nanoscale dissipation and SERS enabled by the combination of optical and electronic transport techniques.  We conclude with a brief discussion of possible further directions.

\section{SAMPLE FABRICATION AND MEASUREMENT}

Devices are fabricated on doped silicon substrate coated with a 200~nm thick thermal SiO$_{2}$ layer.  Electron beam lithography is used to define pads that neck down to form a constriction $\sim$~100~nm wide and 600-800~nm in length.  The structures are metallized via e-beam evaporation of 1~nm of Ti and 15~nm of Au and subsequent liftoff processing.  Organic residue is removed by exposure to oxygen plasma for two minutes.  In samples for use in surface-enhanced Raman spectroscopy, standard surface chemistry procedures are used to self-assemble a monolayer of a molecule of interest on the exposed metal surfaces.  The sample is mounted and wire-bonded in a chip carrier, and loaded into a microscope cryostat.  


Samples are cooled in vacuum ($< 10^{-6}$~mB) to 80~K, and a particular junction is electromigrated in situ via a computer-controlled feedback process to a conductance less than $G_{0} \equiv 2e^{2}/h$.  Electrical measurements are performed using standard lock-in techniques.  A digital-to-analog card is used to source a dc voltage bias, $V_{\mathrm{dc}}$, while a lock-in amplifier sources a small ac bias, $V_{\mathrm{ac}}$ at the frequency $\omega_{0}/2\pi$ typically 2~kHz.  The dc and ac biases are combined with a summing amplifier and applied to one terminal of the junction, while the other terminal is connected to a Keithley 428 current amplifier.  The ac signal is fed into a lock-in amplifier synchronized to $\omega$, measuring the differential conductance, $\partial I/\partial V (V_{\mathrm{dc}})$, and a second channel synchronized to $2\omega$ measuring $(1/4)(\partial^{2}I/\partial V^{2})V_{\mathrm{ac}}^{2}$.

\begin{figure}[h]
  \begin{center}
    \includegraphics[width=10cm,clip]{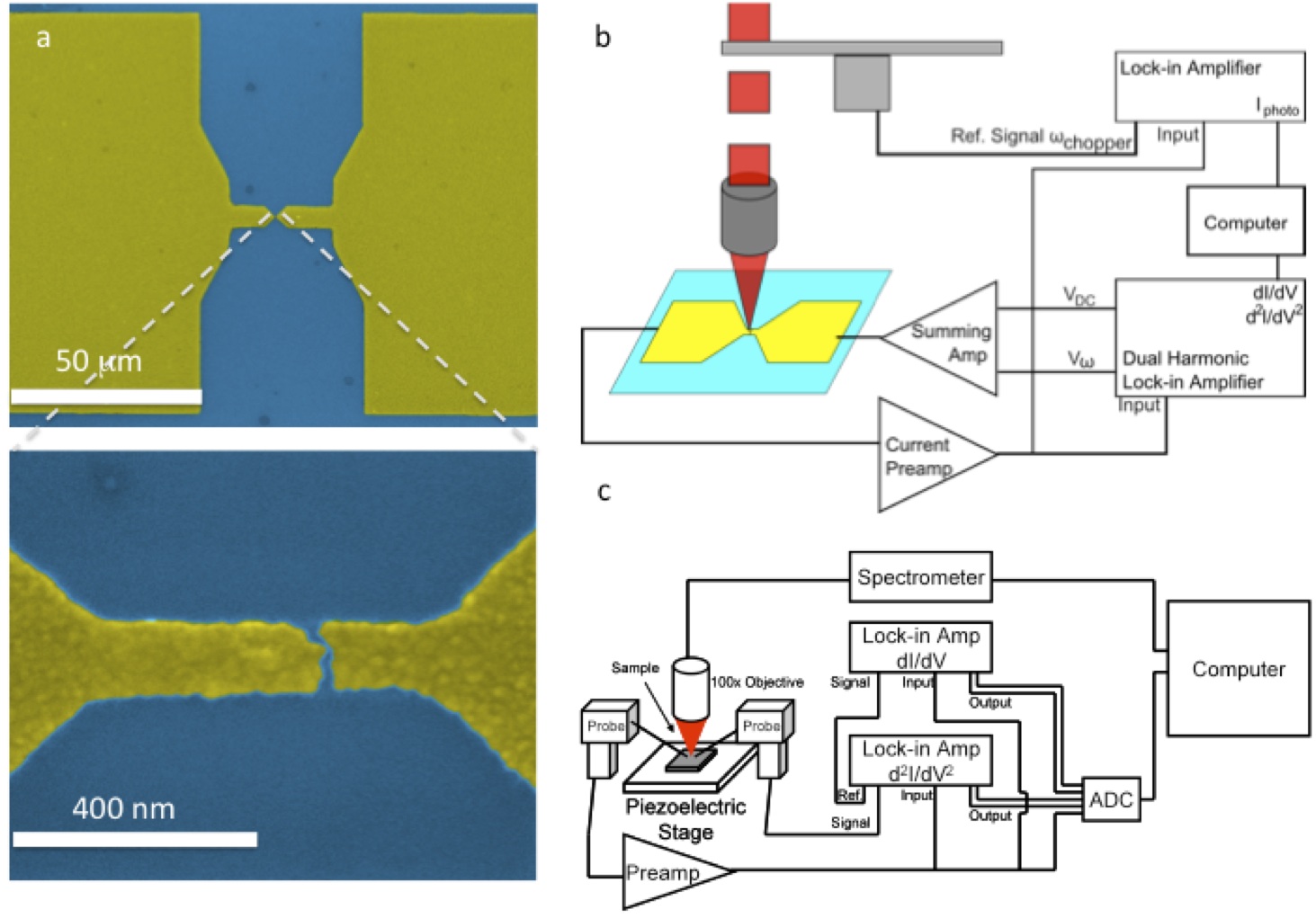}
  \end{center}
  \caption{\label{fig:schem} Experimental configuration.  (a) Electron micrographs showing a typical electromigrated junction, at two magnification levels.  (b) Measurement schematic for optical rectification experiments.  (c) Measurement schematic for simultaneous measurements of electronic transport and Raman scattering.  }
\end{figure}

Sample illumination is performed via a home-made Raman microscope system.  The illumination source is a 785~nm diode laser, spatially filtered, and attenuated to desired intensity levels.  The laser may be rastered over the sample surface with high precision via a piezo-actuated lens.  The beam is focused with a 50$\times$ IR-corrected objective to a gaussian spot size $\sim~1.9~\mu$m, with a maximum intensity at the sample of 23~kW/cm$^{2}$.  The beam may be chopped at a frequency $\omega_{\mathrm{c}}$, enabling lock-in detection of photo-driven processes.  For Raman measurements, spectra are collected using a grating spectrometer with a thermoelectrically cooled silicon CCD detector. 


\section{OPTICAL RECTIFICATION}

We employed electronic transport measurements as a means of assessing the magnitude of the plasmon-enhanced electric fields at the nanogaps.  The concept behind these measurements is \textit{optical rectification}.  Light incident on a metal-vacuum-metal tunnel junction modulates the interelectrode potential difference at optical frequencies by an amount $V_{\mathrm{opt}}$.  In structures excited at their plasmon resonances, one would expect $V_{\mathrm{opt}}$ to be larger than in the off-resonant case.  Likewise, through the ``lightning rod effect'', one would expect a comparatively enhanced $V_{\mathrm{opt}}$ in structures with geometries that favor larger charge displacements.  Suppose that the tunneling conduction process is rapid compared to an optical cycle.   In the presence of a nonlinear tunneling $I-V$ characteristic, classically such a perturbation should contribute to the dc current, by an amount $I_{\mathrm{photo}} = (1/4) (\partial^{2}I/\partial V^{2})V_{\mathrm{opt}}^{2}$, where $(\partial^{2}I/\partial V^{2})$ is the nonlinearity at the particular $V_{\mathrm{dc}}$.  Using low frequency lock-in methods, $(\partial^{2}I/\partial V^{2})$ may be measured at a low frequency as a function of $V_{\mathrm{dc}}$.  

Proportionality between $I_{\mathrm{photo}}(V_{\mathrm{dc}})$ and $(\partial^{2}I/\partial V^{2})(V_{\mathrm{dc}})$ is evidence that the tunneling process at work at long time scales also controls tunneling at optical time scales (particularly if $(\partial^{2}I/\partial V^{2})(V_{\mathrm{dc}})$ has a nontrivial functional form).  Moreover, in the classical rectification model, the precise numerical coefficient of proportionality would be $(1/4)V_{\mathrm{opt}}^{2}$, allowing experimental determination of the optically induced modulation voltage across the tunnel junction.  Since shortly after the invention of the scanning tunneling microscope, measurement of tunneling photocurrent has been suggested as a means of placing a bound on the time scale associated with tunneling.  This approach has been demonstrated unambiguously in STM at microwave frequencies\cite{Tu:2006}.  Visible light experiments in the STM geometry\cite{} have proven more controversial, in part due to confounding effects such as heating-induced variation in tunneling geometry\cite{Tu:2006}.  Electromigrated junctions avoid this issue by having the metal constrained by the underlying substrate.  Measurements in our system show no change in interelectrode separation (inferred from zero-bias conductance) under illumination.

The preceding argument is entirely classical.  Quantum mechanically, we should consider photon-assisted tunneling as the proper framework for this system.  The Tien-Gordon perturbative approach\cite{Tien:1963} is reasonable.  For incident light of angular frequency $\omega$, the Tien-Gordon formalism employs a working parameter $\alpha \equiv eV_{\mathrm{opt}}/\hbar \omega$.  In this approach, $V_{\mathrm{opt}}$ modulates the electronic levels of one electrode relative to the other by an amount $eV_{\mathrm{opt}}$, and tunneling processes are considered that exchange some number of (virtual) photons of energy $\hbar \omega$.  In the limit of small $\alpha$, in the presence of some dc bias $V_{\mathrm{dc}}$, the lowest-order Tien-Gordon correction to the tunneling current is
\begin{equation}
I_{\mathrm{photo}} = I(V_{\mathrm{dc}}, V_{\mathrm{opt}}, \omega)-I(V_{\mathrm{dc}}) = \frac{1}{4}V_{\mathrm{opt}}^{2} \left[ \frac{I(V_{\mathrm{dc}}+ \hbar \omega/e)- 2 I(V_{\mathrm{dc}}) + I(V_{\mathrm{dc}}-\hbar \omega/e)}{(\hbar \omega/e)^{2}}\right].
\label{eq:quantum}
\end{equation}
If the nonlinearity of the dc tunneling conduction $I(V)$ is small from $V_{\mathrm{dc}}-\hbar \omega/e$ to $V_{\mathrm{dc}}+ \hbar \omega/e$, then Eq.~(\ref{eq:quantum}) reduces to the classical rectification expression.  

\begin{figure}[h]
  \begin{center}
    \includegraphics[width=8cm,clip]{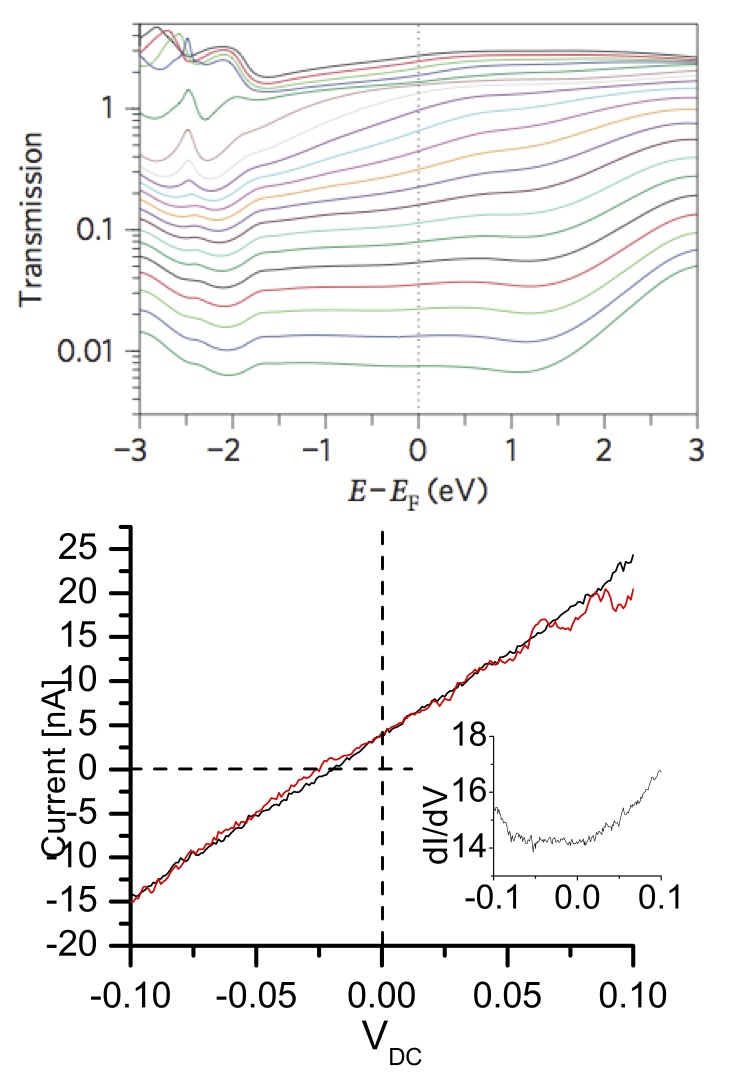}
  \end{center}
  \caption{\label{fig:rectification}Optical rectification.  Top:  Calculated transmission as a function of energy for vacuum tunneling between tip-like Au electrodes.  The different curves correspond to different interelectrode distances, $d$ (in excess of equilibrium bond length), ranging from 2.5 \AA (top) to 7 \AA (bottom) in steps of approximately 0.2 \AA.  The smoothness of transmission over the range of $\pm$~1.5~V around the Fermi level is essential for the validity of classical rectification.  Bottom:  Example data, showing the correspondence of the photocurrent as a function of $V_{\mathrm{dc}}$ and $(1/4)(\partial^{2}I/\partial V^{2})V_{\mathrm{ac}}^{2}$.   }
\end{figure}

Figure~\ref{fig:rectification} (lower panel) shows measurements of the photocurrent for a bare junction (no molecules).  This data was acquired at a substrate temperature of 80~K, with simultaneous measurement of $I$, $\partial I/\partial V$, and $(1/4)(\partial^{2}I/\partial V^{2})V_{\mathrm{ac}}^{2}$ as a function of $V_{\mathrm{dc}}$.  The differential conductance and nonlinearity were measured via lock-in amplifiers using a modest ac drive, $V_{\mathrm{ac}}$, at 2~kHz.  The photocurrent, $I_{\mathrm{photo}}$, was measured using an additional lock-in at $\omega_{\mathrm{c}} = 232$~Hz.  The amplitude $V_{\mathrm{ac}}$ was adjusted deliberately so that the magnitude of  $(1/4)(\partial^{2}I/\partial V^{2})V_{\mathrm{ac}}^{2}$ matched  $I_{\mathrm{photo}}$.  The clear proportionality between the two quantities, including the change in sign at $V_{\mathrm{dc}}\approx -20$~mV, demonstrates the validity of the classical rectification picture in this instance, and $V_{\mathrm{opt}}$ is therefore inferred to be equal to $V_{\mathrm{ac}}$.  Spatial mapping of $I_{\mathrm{photo}}$ as a function of laser position show that only the nanogap region contributes to the photocurrent signal, as expected.  Likewise, for a fixed position, $I_{\mathrm{photo}}$ is directly proportional to the incident optical power. 

Typical values of $V_{\mathrm{opt}}$ found in these measurements are 10-30~mV.  Since the incident photon energy is $\hbar \omega = 1.58$~eV,  $\alpha \approx 0.0063-0.0190$.  Electronic structure calculations (Fig.~\ref{fig:rectification}) based on density functional theory show that the expected transmission as a function of energy across such a gold/vacuum/gold tunnel junction is smooth and relatively featureless over the bias and photon energy range in the experiments.  Thus the analysis in terms of classical rectification is internally consistent.  It is worth noting that the decay of the plasmons in the electrodes should be producing incoherent electron-hole pair excitations, with electron energies as large as $\hbar \omega$ above the Fermi level.  Still, $V_{\mathrm{opt}}$ is the local chemical potential difference produced by the coherent motion of the electrons participating in the nanogap plasmon.

Estimating the enhanced (optical) electric field within the junction requires a length scale, $d$, over which $V_{\mathrm{opt}}$ is dropped.  This is a very subtle issue.  The approach we have taken\cite{Ward:2010} is to estimate $d$ from the interelectrode zero-bias $\partial I/\partial V$.  This is internally consistent, since $V_{\mathrm{opt}}$ is also determined by a tunneling process; the energy difference $e V_{\mathrm{opt}}$ should be the difference in electrochemical potential of the electrodes between the spatial locations of the electronic states at the Fermi level of the source and drain.  When the electrodes are in contact at the single-atom level, the conductance is $G_{0}\equiv 2e^{2}/h$, and electronic structure calculations provide guidance about the decay constant $\beta$ in the expected $\partial I/\partial V \equiv G=G_{0} \exp (-\beta d)$.   The comparatively high conductances of our junctions translate into very small separations, on the {\AA}ngstrom scale.  As a result, the estimated fields within the junctions exceed $10^{8}$~V/m.  For the incident freespace intensities in the experiments, the resulting local field enhancement factors range from several hundred to well over a thousand.  These enhancement factors are consistent with the SERS results that will be discussed \textit{vide infra}.

\begin{figure}[h]
  \begin{center}
    \includegraphics[width=10cm,clip]{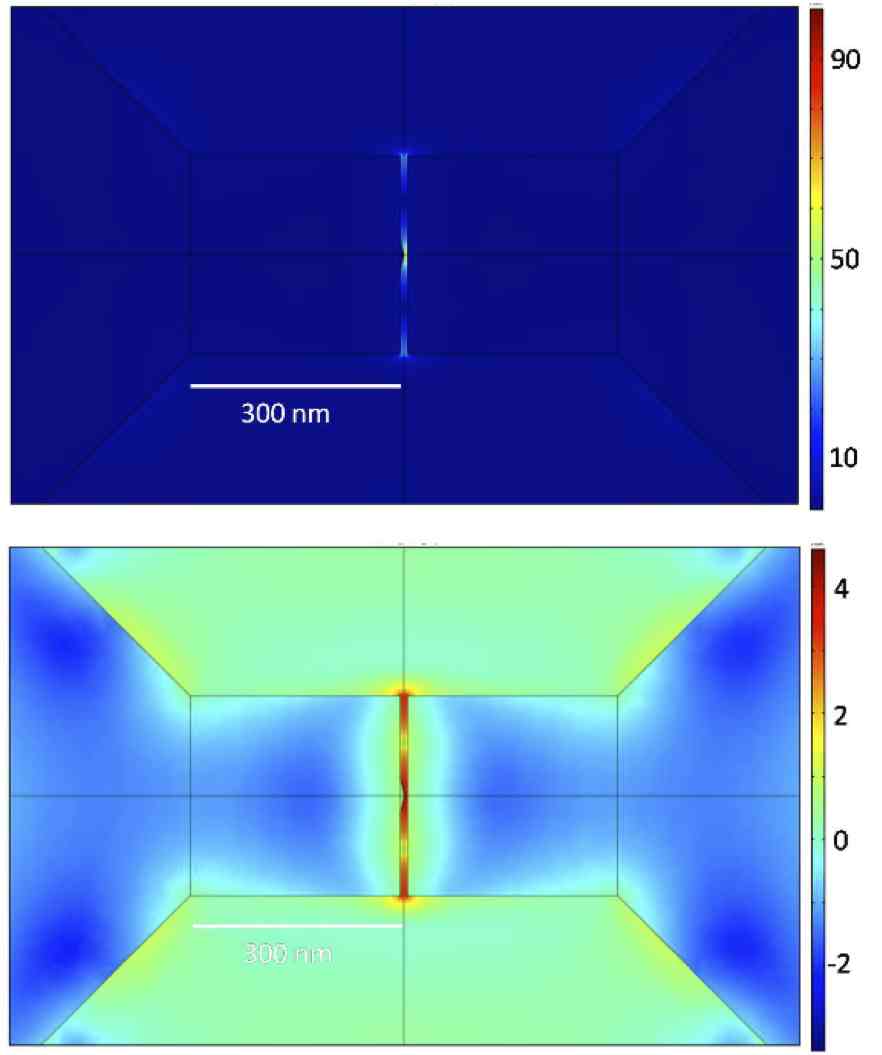}
  \end{center}
  \caption{\label{fig:FE}Finite element model results for a nanoscale gap, illuminated via plane wave (free space wavelength 800 nm) normal to the plane shown, with linear polarization along the interelectrode axis (left-right).  The electrodes are modeled as Au, 15 nm thick, on SiO$_{2}$.  The upper plot has a linear color scale, with electric field normalized to the incident amplitude.  The lower plot has the same normalized data with a logarithmic color scale.  The protrusion on the left electrode in the center increases the field in the gap due to the ``lightning rod'' effect while not drastically altering the plasmon response.  While there is some electric field within the metal, in accord with expectations of the classical skin depth, note that the plasmon response leads to the vast majority of the potential difference between the electrodes being dropped across the interelectrode gap.   }
\end{figure}

However, while the $V_{\mathrm{opt}}$ and $G$ experimental values are unambiguous, the field enhancement values have systematic uncertainties.  A central concern is the combined issue of locality of tunneling and the potential distribution within the metal.  While tunneling is, as discussed, a very local phenomenon, it is important to consider the spatial extent of the initial and final states.  Semiclassically, electrons that tunnel from the source electrode that start out within an inelastic scattering length (several nm at least) of the edge of the sample are unlikely to lose energy prior to tunneling, and therefore probe the instantaneous potential profile extending into the metal.  At low frequencies (used for the $\partial^{2}I/\partial V^{2}$ measurements), the classical screening length in Au is extremely short, a fraction of an atomic diameter.  At optical frequencies, however, there is some field penetration into the metal.  The classical skin depth, which neglects plasmonic effects, in Au at the frequency of interest is $\sim$~25~nm.  Finite-element modeling (Fig.~\ref{fig:FE}) shows that the tip plasmon response does confine the significant majority of the potential drop to the nanoscale vacuum gap between the electrodes.  The consistency between the low frequency nonlinearity data and the photocurrent, observed repeatedly in multiple devices, greatly constrains any differences that may exist in relevant electronic states between the low and optical frequency tunneling processes.

These experiments also highlight the need for treatments of plasmonic response that include the realistic electronic structure.  The finite-element calculations and similar FDTD modeling\cite{} typically use sharp boundaries of dielectric properties that are known to be unphysical at the atomic scale.  In these nanogap structures, with atomically small distances between electrodes, the proper charge distribution within the gap could modify the field distribution from that of the idealized models.  Some theoretical treatments\cite{Liebsch:1993,Liebsch:1993b} have already shown that plasmon excitations involve significant electronic density at {\AA}ngstrom distances outside the metal surface.  A more realistic approach to the electrodynamics of nanostructures that can incorporate electronic structure would be invaluable in trying to understand the ultimate limits possible in plasmonic field enhancements, as well as topics like plasmon-enhanced photochemistry\cite{Watanabe:2006,Brus:2009} and photocatalysis\cite{Chandrasekharan:2000,Awazu:2008}.

\section{SIMULTANEOUS SERS AND TRANSPORT}

As mentioned previously, because of their plasmonic properties, electromigrated junctions have proven to be excellent ``hot spots'' for SERS studies. If a plasmonic structure enhances the local electric field from incident radiation by a factor of $g(\omega)$, then the Raman scattering rate for a molecule in that local field is enhanced from its baseline value by a factor of $|g(\omega)|^{2} |g(\omega')|^{2}$, where $\omega$ is the frequency of incident light, and $\omega'$ is the frequency of the Raman-scattered light.  It has been known for some time\cite{Kneipp:1997,Nie:1997} that sufficient plasmonic enhancement can enable single-molecule Raman sensitivity, though proof of this level of sensitivity is not trivial\cite{Etchegoin:2008}.   Beyond electromagnetic enhancement, it is important to recall that the Raman scattering rate depends on the cross-section, and that the cross-section reflects the full polarization tensor of the molecule plus metal environment.  Due to interactions with the metal, modes may be detectable that would otherwise be Raman-forbidden due to symmetry in the gas phase or solution phase.  Similarly, charge transfer between the metal and molecule can modify the cross-section directly, including possible resonance effects; this SERS contribution is called the ``chemical enhancement''.  

Electromigrated junctions show properties consistent with single-molecule SERS sensitivity.  Following electromigration of a molecular-coated junction, Raman mapping shows a SERS hotspot is apparent at the interelectrode gap, in a high percentage of junctions ($>$~80\% when all of the fabrication, surface chemistry, and electromigration go smoothly).  Positioning the microscope to acquire spectra at that location, time-series of Raman spectra reveal hallmarks of high SERS sensitivity, intensity fluctuations (``blinking'') and spectral diffusion as a function of time.  While it is unlikely that a large ensemble of molecules would show such fluctuations, proving single-molecule sensitivity purely from the statistical properties of Raman spectra of a single analyte alone is exceedlingly challenging\cite{Andersen:2004,LeRu:2006,Domke:2007}.  The origins of this stochastic temporal dependence remain a subject of debate.  

\begin{figure}[h]
  \begin{center}
    \includegraphics[width=8cm,clip]{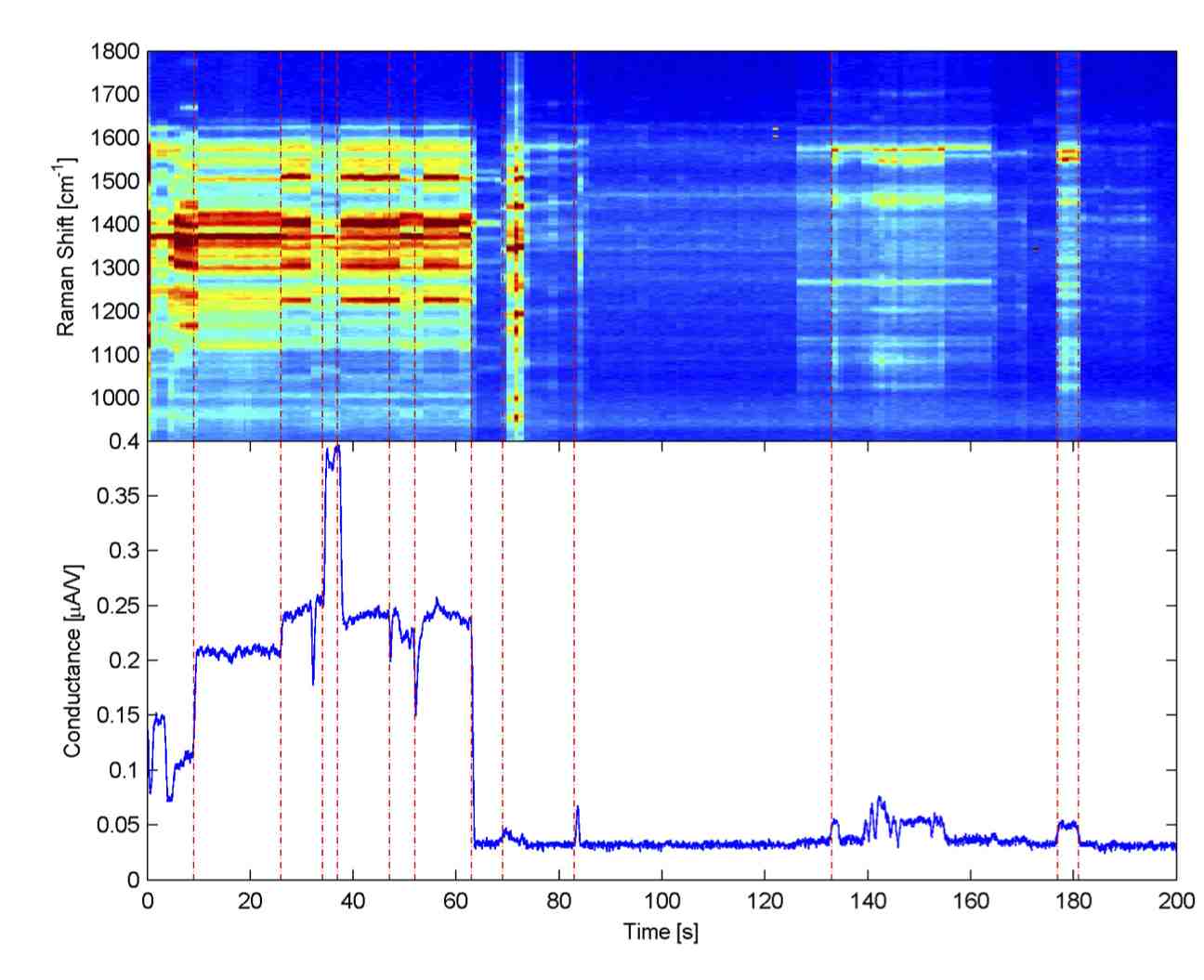}
  \end{center}
  \caption{\label{fig:SERScond}Raman and conductance as a function of time.  Upper panel shows Raman spectra for an oligophenylenevinylene junction measured at 80~K, from 900~cm$^{-1}$ to 1800~cm$^{-1}$.  Dark blue is zero counts, while dark red is 1000 counts per second. The lower panel shows simultaneously measured differential conductance at zero bias.  Dashed lines have been added to highlight particular changes in the conductance; each corresponds with a change in the Raman spectrum.  Tunneling conductance is known to be dominated by a molecular-scale volume. }
\end{figure}

Simultaneously acquired interelectrode conductance measurements provide crucial additional information.  In junctions with measurable tunneling conductance, we frequently observe correlations in the time dependence of the conductance and the time dependence of the Raman spectra.  Figure~\ref{fig:SERScond} shows an example of a typical case, acquired at a substrate temperature of 80~K.  The ac excitation used to measure conductance was 10~mV(rms), and the integration time for Raman acquisition was 1 s.  Discrete changes in the Raman spectrum correlate with changes in the interelectrode tunneling conductance.  Past FDTD calculations, \textit{in situ} measurements of Raman during electromigration\cite{Ward:2008}, and the rectification measurements\cite{Ward:2010} discussed here all confirm that small changes in interelectrode conductance when $G \ll G_{0}$ cause changes in overall field enhancement, but not complex changes in the Raman spectrum structure.  Since the tunneling conductance is sensitive to a molecular-scale volume in the interelectrode gap, the observed correlatons imply that the Raman spectrum originates with a molecule or molecules in that gap.   

Note that the idealized concept of a single molecule, fully extended along the interelectrode direction, neatly bridging the two electrodes, is very unlikely to be the case in practice in these kinds of electromigrated junctions.  It is far more likely that the molecule is more strongly coupled to one electrode than the other.  Likewise, only some portion of the molecule may lie within the region probed by tunneling current, and there can be a significant contribution to the current from direct metal-metal tunneling.  However, the observed correlations between Raman and conductance imply that the dominant source of the Raman scattering is some molecule or molecules with an electronic interaction with the tunneling electrons.

Given that some fraction of the measured tunneling current is interacting with a Raman-active molecule in such junctions, it is natural to consider using the Raman emission as a means of probing the transport process.  Transport experiments through molecular junctions have demonstrated the importance of electron-vibrational interactions.  For example, at sufficiently high source-drain bias, it is possible to have sequential tunneling transport through a vibrationally excited state of the next molecular electronic level.  This has been observed in fullerenes\cite{Park:2000,Park:2003} in single-molecule transistor experiments, and in scanning tunneling spectroscopy (STS) studies of copper phthallocyanine\cite{Qiu:2004}, for example.    Off resonance, inelastic cotunneling via vibrationally excited states of the molecule is the mechanism behind inelastic tunneling spectroscopy (IETS)\cite{Lambe:1968}.  IETS in single-molecule junctions has been demonstrated with STS\cite{Stipe:1998} and in single-molecule transistors\cite{Yu:2004,Song:2009}.  If these processes are sufficiently active, and the relaxation rate of local vibrational modes to the substrate's bulk phonons is sufficiently slow, then current flow should perturb the vibrational population away from thermal equilibrium.  For a particular Raman-active vibrational mode, population of the vibrational excited state should be detectable through the anti-Stokes Raman process.  Intriguing experiments along these lines in ensembles of molecules have been reported\cite{Ioffe:2008}.

At a particular temperature $T$, if the Raman process is a weak perturbation on the vibrational population, the ratio of anti-Stokes to Stokes Raman emission for a particular mode should be:
\begin{equation}
\frac{I^{\mathrm{AS}}}{I^{\mathrm{S}}} = \left( \frac{\sigma_{\mathrm{AS}}}{\sigma_{\mathrm{S}}}\right) \left(\frac{g(\omega_{\mathrm{AS}})}{g(\omega_{\mathrm{S}})}\right)^{2} \left( \frac{\omega_{\mathrm{AS}}}{\omega_{\mathrm{S}}}\right)^{4} \exp\left(-\frac{ \hbar (\delta \omega)}{k_{\mathrm{B}}T}\right).
\label{eq:Teff}
\end{equation}
Here $\sigma_{\mathrm{AS}}$ ($\sigma_{\mathrm{S}}$) is the anti-Stokes (Stokes) Raman cross-section, $\omega_{\mathrm{AS}}$ ($\omega_{\mathrm{S}}$) is the frequency of the anti-Stokes (Stokes) emission, $g$ is the frequency-dependent electromagnetic enhancement factor from the plasmonic environment, $\delta \omega$ is the Raman shift of the particular vibrational mode, and $k_{\mathrm{B}}$ is Boltzmann's constant.   Even out of thermal equilibrium, if we are concerned only with a vibrational ground state and a single excited vibrational level, then we can always use Eq.~(\ref{eq:Teff}) to infer an \textit{effective} temperature for a given vibrational mode.  

The Raman process itself drives the vibrational population out of thermal equilibrium.  This is a mechanism of optical pumping in SERS, in which the anti-Stokes signal scales quadratically with incident intensity, while the Stokes signal scales linearly as usual\cite{Galloway:2009}.  The efficacy of this process again depends on the relative rates of Stokes Raman scattering (populating the vibrationally excited state) and relaxation to the bulk phonons (returning the system to the vibrational ground state).  An additional mechanism at work in SERS is the interaction of the molecule with hot electrons in the metal.  The hot electrons and holes created by optical absorption and plasmon decay can also couple to vibrational modes.  This, too, would lead to a quadratic dependence of anti-Stokes intensity on incident power.

\begin{figure}[h]
  \begin{center}
    \includegraphics[width=10cm,clip]{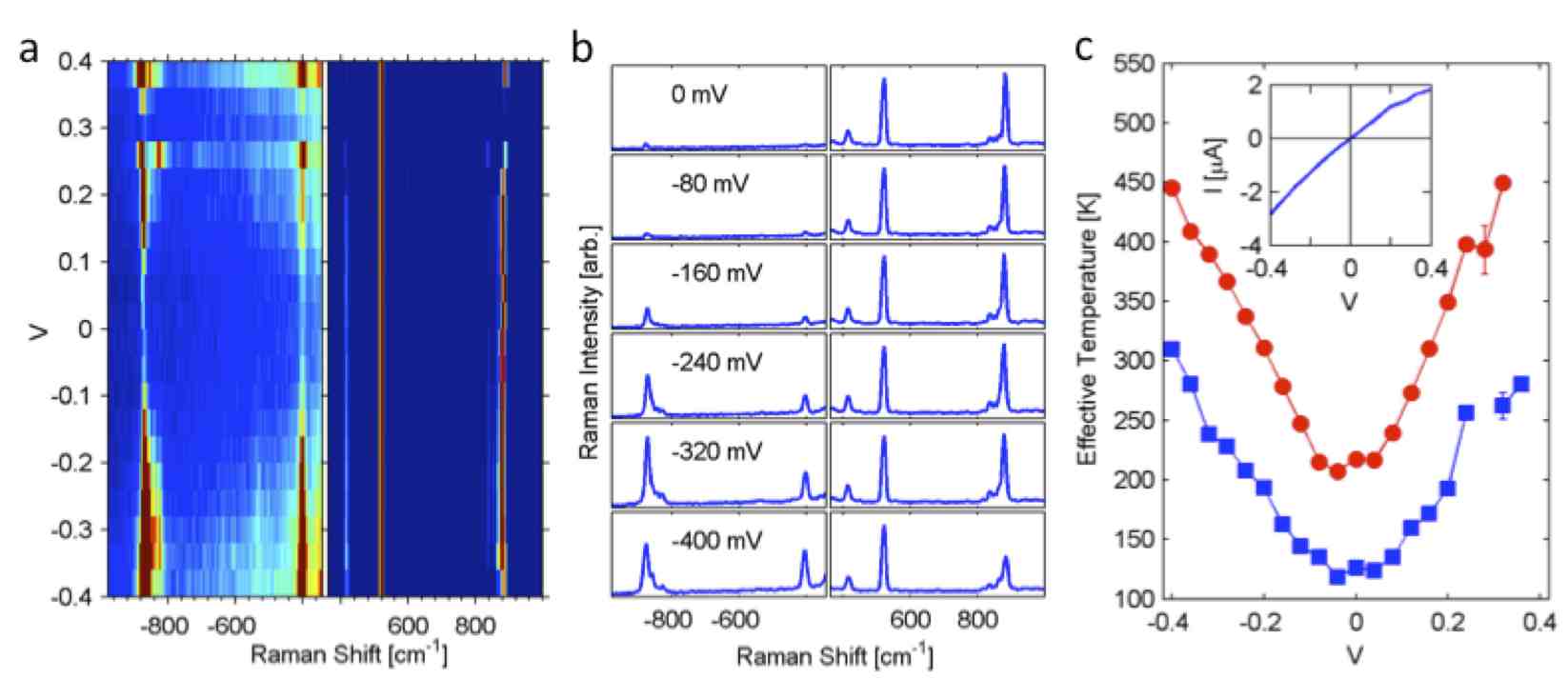}
  \end{center}
  \caption{\label{fig:biasdata} Raman spectra as a function of bias conditions for an electromigrated junction containing an oligophenylenevinylene molecule\protect{\cite{Ward:2011}}.  (a) At left (right), anti-Stokes (Stokes) Raman as a function of $V_{\mathrm{dc}}$ applied across the junction.  Color scale:  Blue indicates 10 (2500) counts, while red indicates 100 (8000) counts per 1 s integration for anti-Stokes (Stokes) emission.  The strong peak at 520~cm$^{-1}$ on the Stokes side is the Raman emission from the Si substrate.  Note the strong increase of anti-Stokes signal at high bias.  (b) Raman spectra at particular bias voltages.  At left (right), anti-Stokes (Stokes) Raman as a function of wavenumber, with full scale corresponding to 1200 (18000) counts per second for anti-Stokes (Stokes) data. (c) Vibrational effective temperatures (inferred from Eq.~(\ref{eq:Teff}) assuming the cross-section and plasmon enhancement ratios are one) as a function of bias.  Red (blue) points correspond to the 880 (410)~cm$^{-1}$ mode Inset:  Simultaneously acquired current-voltage characteristics.  }
\end{figure}

 The cross-sections and plasmonic dispersion in Eq.~(\ref{eq:Teff}) are very difficult to determine in specific nanostructures.  However, even without that knowledge it is possible to use the effect of bias on the anti-Stokes/Stokes SERS emission to learn about vibrational energy distributions on the single molecule level.  Figure~\ref{fig:biasdata} shows an example of bias-driven vibrational excitation, including some amount of optical pumping, for a junction made with amine-terminated oligophenylenevinylene (OPV3) molecules assembled on the electrodes .  The zero-bias dc conductance of this junction is $\sim 0.06~G_{0}$, suggesting that there is a significant contribution to the conduction from direct metal-metal tunneling.    

Even at zero bias, there is some detectable anti-Stokes SERS emission.  This is believed to result from optical pumping, by the following argument.  Assuming the unknown ratios in Eq.~(\ref{eq:Teff}) to be one, then the expected anti-Stokes/Stokes ratio for the 880~cm$^{-1}$ mode should be $2 \times 10^{-7}$ for an equilibrium temperature of 80 K.  Instead, the ratio is closer to $5 \times 10^{-3}$, consistent with a population equivalent to a zero-bias effective temperature of $\sim$~220~K.  This difference is too large to be a systematic issue with the cross-section or plasmonic enhancement ratios.   Given that the local substrate temperature (based on the silicon Raman signal) is consistent with 80~K, it is reasonable to conclude that the elevated vibrational population at zero bias is due to optical pumping.      

As bias is increased, in either polarity, the anti-Stokes intensity increases.  Again characterized by an effective temperature, this rise in excited vibrational population is clear in Fig.~\ref{fig:biasdata}a.  For the 880~cm$^{-1}$ mode, significant changes in the vibrational population begin to appear when the bias exceeds $V_{\mathrm{DC}} \approx 0.11$~V, the vibrational mode energy.  Resolution is not sufficient to address this in the lower energy mode.   This sort of thresholding has been observed in other junctions\cite{Ward:2011}, and is consistent with the notion that this is current-driven vibrational pumping.  The alternative, that the current flow locally heats the metal which in turn excites the vibrations, would not be expected to have any particular thresholding, and would be expected to affect all vibrational modes equally, in contrast to what is observed.

Note that there is a discontinuity in the data at a positive bias near 300~mV.  This is a blinking event, with a corresponding change in conductance.  After the time required to take two voltage data points, the junction returns to its previous state.  A major challenge in studying bias-driven changes in the Raman is the balance between the competing requirements that conduction and Raman be correlated on the one hand, and that the junction be stable (non-blinking) long enough to acquire the bias-dependent data on the other.

In addition to the current-driven changes in discrete Raman modes on the anti-Stokes side, there is also a continuum Raman emission at low wavenumbers that is enhanced as bias is increased.  Continuum emission at low Raman shift in SERS systems has previously been identified\cite{Moskovits:2005,Otto:2006,Jiang:2003} as Raman scattering by the electrons in the plasmonically active metal.  No complete theory of this effect currently exists\cite{Mahajan:2010}.  There are kinematic restrictions on electronic Raman scattering, and disorder/impurity scattering (including adsorbed molecules) affect the conservation of electronic crystal momentum.  Electronic anti-Stokes emission requires recombination of an electron and a hole with an energy difference equal to the Raman shift.  An increase in the anti-Stokes Raman continuum at high biases suggests that the current flow leads to more electrons available above the Fermi energy for such a process - heating of the electronic distribution.  This is discussed more quantitatively elsewhere\cite{Ward:2011}, though there is definitely a need for an improved theoretical treatment of such processes.

\section{Discussion and future directions}

Combined optical and transport measurements in these plasmonic nanostructures opens up many possibilities.  Further work in this direction can have an impact on our understanding of nanoscale electronic transport and fundamental issues of dissipation; the properties and uses of ultrasmall plasmonic structures; and the unsolved physical chemistry questions relevant to SERS and other surface-enhanced spectroscopies.

From the perspective of nanoscale electronic transport, latent in the Raman and conductance data of Fig.~\ref{fig:SERScond} is a great deal of information about the precise configuration of the molecule at the junctions.  Recent advances in computational tools, particularly for determining polarizability tensors with realistic electronic structure of molecules and surfaces, show great promise in this direction\cite{Zayak:2011}.  SERS measurements in these junctions are one of the few ways that one can imagine accessing detailed configurational information.  

Similarly, SERS in these structures is a means of examining the vibrational and even electronic distribution functions, and how these are affected by the flow of current.  Beyond equilbrium SERS, the apparent vibrational pumping is determined by the rate of current flow, the energy of the flowing electrons, the coupling of those electrons to local modes, and the coupling of those modes to the underlying substrate.  The hope is that such studies will allow the determination of parameters like the electron-vibrational coupling, a quantity that is otherwise difficult to measure experimentally \textit{in situ}.  A number of theoretical investigations of heating in current-carrying junctions have been performed (see Dubi and di Ventra and references therein\cite{Dubi:2011}).  These measurements are a comparatively direct experimental means of assessing the relevant vibrational and electronic degrees of freedom.  Time-resolved optical and electronic studies would be even more revealing, potentially allowing the examination of the approach to steady state, though such experiments would require significant further experimental development. 

The plasmonic properties of the nanojunctions also bear further examination.  The rough length scale for the confinement of the enhanced local electric field in these gaps is approximately the geometric mean of the interelectrode spacing and the radius of curvature of the tips of the electrodes.  While calculations\cite{Ward:2007} show resonances in Au junctions extending from 700~nm  well into the infrared, there has yet to be a systematic experimental examination of the plasmon properties beyond $\sim$~900~nm.  Of central interest is the junction-to-junction variability in the resonances, their peak wavelengths, amplitudes, and quality factors.  These should depend in detail on how the real nanogap breaks geometric symmetries about the gap center.  The surprisingly good reproducibility of strong SERS under 785~nm illumination raises interesting questions about the sensitivity of these modes to detailed electrode morphology. 

The use of such nanogaps for other plasmonic applications is an exciting prospect.  While the typical quality factor of such a plasmon resonance is comparatively poor ($\sim$10-20), the effective mode volume is extremely small.  Other plasmonic structures have been considered for similar reasons for interesting Purcell effects\cite{Schuller:2010,Vesseur:2010}, and novel nanolasers based on plasmon resonances have recently been demonstrated\cite{Noginov:2009,Oulton:2009}.  Future investigations to integrate nanogap plasmons to make novel emitters and absorbers will be interesting.

Finally, these types of experiments raise the hope of addressing some persistent problems in the physical chemistry of surface-enhanced spectroscopies.  The physical mechanisms behind the intensity fluctuations and spectral diffusion commonly observed at single-molecule sensitivities remain elusive.  The nanogap architecture allows examination of some candidate ideas.  For example, small systematic changes are apparent in the Stokes Raman shift in junctions under a dc voltage bias, particularly when conductance is immeasurably small.  These suggest that Raman Stark physics\cite{Lambert:1984} due to nearby charge traps or adsorbed impurities can modify Raman shifts by amounts comparable to those observed in spectral diffusion measurments.  Similarly, the origins of blinking in SERS have been studied for some time, with several candiate explanations put forward\cite{Wang:2005}.  Studies in these kinds of individual hotspots should enable testing of ideas concerning molecule-metal charge transfer, especially since the underlying substrate may be used as a gate electrode to tune molecular level alignment relative to the source and drain.  Such studies should also be able to examine the nature of chemical enhancement in individual junctions, without the need for electrolytes.  Nanoscale gaps for optical measurements are a powerful addition to the experimental toolkit, in electronic transport, plasmonics, and physical chemistry.

\section*{ACKNOWLEDGMENTS}
D.N. and D.R.W. acknowledge support by the Robert A. Welch Foundation (grant C-1636) and the Lockheed Martin Advanced Nanotechnology Center of Excellence at Rice (LANCER). D.N. and D.R.W. acknowledge valuable conversations with M. Di Ventra, M.A. Ratner and A. Nitzan.  F.H. and J.C.C. acknowledge support from the CFN, the DFG SPP 1243, the Baden-W{\"u}rttemberg Stiftung within the Network of Excellence 'Functional Nanostructures', and the Spanish Ministry of Science and Innovation (Ministerio de Ciencia e Innovacion) (grant FIS2008-04209). F.P. acknowledges funding from a Young Investigator Group.


\end{document}